\documentclass[12pt]{article}
\usepackage{amsmath}

\newtheorem{theorem}{Theorem}
\newtheorem{acknowledgement}[theorem]{Acknowledgement}

\input{tcilatex}
\begin{document}

\begin{titlepage}
\vspace{.3cm} \vspace{1cm}
\begin{center}
\baselineskip=16pt \centerline{\Large\bf Gravity with de Sitter and Unitary Tangent Groups } \vspace{2truecm} \centerline{\large\bf Ali H.
Chamseddine$^{1,3}$\ , \ Viatcheslav Mukhanov$^{2,4}$\ \ } \vspace{.5truecm}
\emph{\centerline{$^{1}$Physics Department, American University of Beirut, Lebanon}}
\emph{\centerline{$^{2}$Theoretical Physics, Ludwig Maxmillians University,Theresienstr. 37, 80333 Munich, Germany }}
\emph{\centerline{$^{3}$I.H.E.S. F-91440 Bures-sur-Yvette, France}}
\emph{\centerline{$^{4}$Department of Physics, New York University, NY 10003, USA}}
\end{center}
\vspace{2cm}
\begin{center}
{\bf Abstract}
\end{center}
Einstein Gravity can be formulated as a gauge theory with the tangent space respecting  the Lorentz symmetry. In this paper we show that the dimension of the tangent space can be
larger than the dimension of the manifold and  by requiring the
invariance of the theory with respect to 5d Lorentz group (de Sitter group) Einstein theory is reproduced unambiguously. The other possibility is to have unitary symmetry on a complex tangent space of the same dimension as the manifold. In this case
the resultant theory is Einstein-Strauss Hermitian gravity. The tangent group is important for matter couplings. We show that in the de Sitter case the 4 dimensional space time vector and scalar are naturally unified by a hidden symmetry being components of a 5d vector in the tangent space.
With a de Sitter tangent group spinors can exist only when they are made complex or taken in doublets in a way similar to N=2 supersymmetry.
\end{titlepage}

\section{\protect\bigskip Introduction}

The experimental evidence that Lorentz symmetry is preserved for effective
four-dimensional theories is overwhelming. In curved space-time this Lorentz
symmetry is realized as a local symmetry of the tangent manifold \cite%
{utiyama} \cite{Kibble}. Moreover, to incorporate spinors in general
relativity, we are forced to consider this local symmetry because there are
no spinor representations of the diffeomorphism group. Usually the dimension
of the tangent space is taken to be equal to the dimension of the curved
manifold and then the Lorentz symmetry is simply a manifestation of the
equivalence principle, which is valid in torsion-free theories. General
relativity could then be formulated as a gauge theory of the Lorentz group
where the gauge fields are the spin-connection. In reality one can search
for all possible tangent groups in $d$-dimensional space-time \cite{Wein}.
In this paper we will investigate whether it is possible to have a larger
group of symmetry in the tangent space and still unambiguously reproduce
general relativity. We will show in section 2, that this is indeed possible
by taking the tangent space to be real with de Sitter group symmetry. The de
Sitter gauge invariant action which is linear in curvature is shown to be
identical to Einstein gravity, provided that metricity condition is imposed
on the spin and affine connections. In section 3 we consider matter
interactions of gravity with the de Sitter group as the tangent group. We
then, in section 4, consider a complex tangent space and show that the
relevant symmetry in this case is the unitary symmetry. The resultant theory
is the Einstein-Strauss theory. Section 5 is the conclusion. An appendix
treats the special limit of Poincare symmetry, and examines the relation of
our new formalism in three dimensions with Witten's formulation of
Chern-Simons gravity.

\section{Gravity with de Sitter tangent group}

Let us begin with a $d$-dimensional manifold and assume that at every point
of this manifold there is a real $N$-dimensional tangent space spanned by
linearly independent vectors $\mathbf{v}_{A}$, where $A=1,2...N.$ Assuming
that $d\leq N$, the coordinate basis vectors $\mathbf{e}_{\alpha }\equiv
\partial /\partial x^{\alpha },$ where $\alpha =1,2...d,$ span $d$%
-dimensional space. Next we define the scalar product in the tangent space
and take the vectors $\mathbf{v}_{A}$ to be orthonormal\footnote{%
We use the notation and methods of Misner, Thorne and Wheeler (\cite{MTW}),
in particular Chapters 9 and 10.} 
\begin{equation}
\mathbf{v}_{A}\cdot \mathbf{v}_{B}=\eta _{AB}.  \label{1}
\end{equation}%
where $\eta _{AB}$ is Minkowski matrix. The Lorentz transformations 
\begin{equation}
\mathbf{\tilde{v}}_{A}=\Lambda _{A}^{\hspace{0.05in}B}\mathbf{v}_{B},\text{
\ \ \ \ \ \ }\Lambda _{A}^{\hspace{0.05in}C}\eta _{CD}\Lambda _{A}^{\hspace{%
0.05in}D}=\eta _{AB}\text{\ }  \label{1a}
\end{equation}%
preserve the orthogonality of the vielbein, $\mathbf{\tilde{v}}_{A}\cdot 
\mathbf{\tilde{v}}_{B}=\eta _{AB}.$ The scalar product of coordinate basis
vectors then induces the metric in $d$-dimensional manifold%
\begin{equation}
\mathbf{e}_{\alpha }\cdot \mathbf{e}_{\beta }=g_{\alpha \beta }(x^{\gamma }).
\label{2}
\end{equation}%
Expanding $\mathbf{e}_{\alpha }$ in $\mathbf{v}_{A}$-basis%
\begin{equation}
\mathbf{e}_{\alpha }=e_{\alpha }^{B}\mathbf{v}_{B},  \label{3}
\end{equation}%
and substituting in (\ref{2}) we obtain the following expression for the
metric $g_{\alpha \beta }$ 
\begin{equation}
g_{\alpha \beta }=e_{\alpha }^{A}e_{\beta }^{B}\eta _{AB},  \label{4}
\end{equation}%
in terms of components. Tangent space indices are raised and lowered with
the Minkowski metric, thus%
\begin{equation}
e_{A\alpha }=\eta _{AB}e_{\alpha }^{B}=\left( \mathbf{v}_{A}\cdot \mathbf{e}%
_{\alpha }\right) ,  \label{4a}
\end{equation}%
and $\eta ^{AB}$ is inverse to Minkowski matrix $\eta _{AB}.$ Next we
consider parallel transport on the manifold relating vectors in
\textquotedblleft nearby\textquotedblright\ tangent spaces. The affine and
spin connections determining the rules for parallel transport of the
coordinate basis vectors and vielbein are defined via%
\begin{equation}
\mathbf{\nabla }_{\mathbf{e}_{\beta }}\mathbf{e}_{\alpha }\equiv \mathbf{%
\nabla }_{\beta }\mathbf{e}_{\alpha }=\Gamma _{\alpha \beta }^{\nu }\mathbf{e%
}_{\nu },\ \ \mathbf{\nabla }_{\beta }\mathbf{v}_{A}=-\omega _{\beta
A}^{\quad \hspace{0.03in}B}\mathbf{v}_{B},  \label{5a}
\end{equation}%
where $\mathbf{\nabla }_{\beta }$ is the derivative defining the rate of
change of vectors along a basis vector $\mathbf{e}_{\beta }$. When applied
to a scalar function $f$ \ this derivative acts as a partial derivative with
respect to the appropriate coordinates, that is, $\mathbf{\nabla }_{\beta
}f=\partial f/\partial x^{\beta }$. Notice that $\eta _{AB}$ and $g_{\alpha
\beta }$ as defined in (\ref{1}) and (\ref{2}) are the sets of scalar
functions and, hence, $\mathbf{\nabla }_{\beta }\eta _{AB}=0,$ $\mathbf{%
\nabla }_{\gamma }g_{\alpha \beta }=\partial g_{\alpha \beta }/\partial
x^{\gamma }\equiv \partial _{\gamma }g_{\alpha \beta }$.

Given $\eta _{AB},$ $g_{\alpha \beta }$ and $e_{\alpha }^{A}$ we derive the
consistency (metricity) conditions for the connections by taking derivative
of equations (\ref{1}), (\ref{2}) and (\ref{4a}). In particular, we obtain%
\begin{equation}
\left( \mathbf{\nabla }_{\alpha }\mathbf{v}_{A}\right) \cdot \mathbf{v}_{B}+%
\mathbf{v}_{A}\cdot \left( \mathbf{\nabla }_{\alpha }\mathbf{v}_{B}\right)
=-\omega _{\alpha AB}^{\quad }-\omega _{\alpha BA}^{\quad }=\mathbf{\nabla }%
_{\alpha }\eta _{AB}=0,  \label{6}
\end{equation}%
that is, the spin connection should be antisymmetric with respect to tangent
space indices, $\omega _{\alpha AB}^{\quad }=-\omega _{\alpha BA}^{\quad }.$
Applying the derivative $\mathbf{\nabla }_{\gamma }$ to (\ref{2}) gives%
\begin{equation}
\Gamma _{\alpha \gamma }^{\nu }g_{\nu \beta }+\Gamma _{\beta \gamma }^{\nu
}g_{\alpha \nu }=\partial _{\gamma }g_{\alpha \beta }  \label{7}
\end{equation}%
Assuming that torsion is absent, $\Gamma _{\alpha \beta }^{\nu }=\Gamma
_{\beta \alpha }^{\nu }$, these equations are solved unambiguously, giving
the well known result%
\begin{equation}
\Gamma _{\alpha \beta }^{\gamma }=\frac{1}{2}g^{\gamma \sigma }\left(
g_{\alpha \sigma ,\beta }+g_{\sigma \beta ,\alpha }-g_{\alpha \beta ,\sigma
}\right) ,  \label{8}
\end{equation}%
where $g^{\gamma \sigma }$ is inverse to $g_{\alpha \beta },$ that is, $%
g^{\alpha \sigma }g_{\sigma \beta }=\delta _{\beta }^{\alpha }.$ We would
like to stress that affine connections are determined unambiguously
irrespective of the group of tangent space. Finally, from (\ref{4a}) we
obtain%
\begin{equation}
\partial _{\beta }e_{A\alpha }=-\omega _{\beta A}^{\quad \hspace{0.03in}%
B}e_{B\alpha }+\Gamma _{\alpha \beta }^{\nu }e_{A\nu }.  \label{9}
\end{equation}%
Let us find when these equations can unambiguously be solved for $\omega
_{\beta B}^{\quad \ A}$ in terms of the soldering form $e_{\alpha }^{B}$ and
metric $g_{\alpha \beta }.$ The total number of components of $e_{\alpha
}^{B}$ is $Nd$. Given a metric $g_{\alpha \beta },$ whose derivatives
determine $\Gamma _{\alpha \beta }^{\nu }$ via (\ref{8}), and hence impose $%
\frac{1}{2}d^{2}\left( d+1\right) $ constraints on $\partial _{\beta
}e_{A\alpha }$, leaves us with $d\left( Nd-\frac{1}{2}d\left( d+1\right)
\right) $ \textit{independent} equations (\ref{9}) to determine $\frac{1}{2}%
dN\left( N-1\right) $ antisymmetric spin connections $\omega _{\beta
AB}^{\quad }.$ Note that for any $N$ and $d$ \ the number of equations can
never exceed the number of independent $\omega _{\beta AB}^{\quad }$ to be
determined, and hence for any dimension of tangent space the system of
equations is not overdetermined. However, the spin connection is
unambiguously determined only if the number of equations is equal to the
number of its unknown components:%
\begin{equation*}
d\left( Nd-\frac{1}{2}d\left( d+1\right) \right) =\frac{1}{2}dN\left(
N-1\right) .
\end{equation*}%
The only solutions of this equation are $N=d$ \ and $N=d+1.$ The first case
is well known and thus we shall concentrate on the second case which
corresponds to the larger symmetry group $SO(1,d)$ of the tangent space. In
the case of a four-dimensional manifold the tangent space is five
dimensional. The metric in 5d tangent space can then be taken either to be $%
\eta _{AB}=\mathrm{diag}\left( 1,-1,-1,-1,-1\right) $ or $\eta _{AB}=\mathrm{%
diag}\left( 1,1,-1,-1,-1\right) $. In the first case the gauge group is 5d
Lorentz group $SO(1,4)$ which is also the group of symmetry of 4d de Sitter
space (de Sitter group), while in the second case the group is $SO(2,3)$
(the group of symmetry of 4d anti de Sitter space). For definiteness and
from \ here on, we consider these cases only. Note that although the
consistency equations do not lead to any contradiction for an arbitrary
dimension of tangent space the connections are entirely determined by the
soldering form only if $N=d$ or $N=d+1.$ Otherwise the spin connection is
not unambiguously determined by the fundamental soldering form and the
theory is not well defined.

In order to construct gauge invariant Lagrangians we need to define%
\begin{equation}
e_{A}^{\alpha }=g^{\alpha \gamma }e_{\gamma A}=g^{\alpha \gamma }\eta
_{AB}e_{\gamma }^{B}.  \label{10}
\end{equation}%
Rewritten in terms of $e_{A}^{\alpha },$ equation (\ref{9}) becomes 
\begin{equation}
\partial _{\beta }e_{A}^{\alpha }=-\omega _{\beta A}^{\quad \hspace{0.03in}%
B}e_{B}^{\alpha }-\Gamma _{\nu \beta }^{\alpha }e_{A}^{\nu }.  \label{11}
\end{equation}%
The soldering form $e_{A}^{\alpha }$ is inverse to $e_{\beta }^{B}$ only if
the of dimension of the tangent space and the dimension of the manifold
match. In case of a de Sitter tangent group contraction over tangent space
indices gives%
\begin{equation}
e_{A}^{\alpha }e_{\beta }^{A}=g^{\alpha \gamma }\eta _{AB}e_{\gamma
}^{B}e_{\beta }^{A}=g^{\alpha \gamma }g_{\gamma \beta }=\delta _{\beta
}^{\alpha },  \label{12}
\end{equation}%
however, contraction over space-time indices gives 
\begin{equation}
e_{A}^{\alpha }e_{\alpha }^{B}\neq \delta _{B}^{A}.
\end{equation}%
To prove this, let us introduce the unit vector $\mathbf{n}$ orthogonal to
all $\mathbf{e}_{\alpha },$ that is, $\mathbf{n}\cdot \mathbf{e}_{\alpha }=0$
and $\mathbf{n}\cdot \mathbf{n=}\varepsilon ,$ where $\varepsilon =-1$ or $%
+1 $ for de Sitter and anti de Sitter groups correspondingly. The vectors $%
\mathbf{n}$ and $\mathbf{e}_{\alpha }$ form a complete basis in tangent
space and therefore 
\begin{equation}
\mathbf{v}_{A}=v_{A}^{\alpha }\mathbf{e}_{\alpha }+n_{A}\mathbf{n.}
\label{13}
\end{equation}%
Taking into account (\ref{4a}) we have%
\begin{equation}
v_{A}^{\alpha }=g^{\alpha \gamma }\left( \mathbf{v}_{A}\cdot \mathbf{e}%
_{\gamma }\right) =g^{\alpha \gamma }\eta _{AB}e_{\gamma }^{B}=e_{A}^{\alpha
},  \label{14}
\end{equation}%
that is, the soldering form $e_{A}^{\alpha }$ coincides with the coefficient 
$v_{A}^{\alpha }$ in expansion (\ref{13}). Taking this into account one gets 
\begin{equation}
\eta _{AB}=\mathbf{v}_{A}\cdot \mathbf{v}_{B}=v_{A}^{\alpha }v_{B}^{\beta
}g_{\alpha \beta }+\varepsilon n_{A}n_{B}=e_{A}^{\alpha }e_{\alpha
B}+\varepsilon n_{A}n_{B},  \label{15}
\end{equation}%
or after rasing the tangent space index we obtain%
\begin{equation}
e_{A}^{\alpha }e_{\alpha }^{B}=\delta _{A}^{B}-\varepsilon n_{A}n^{B}\equiv
P_{B}^{A}  \label{16}
\end{equation}%
where $P_{B}^{A}$ is a projection operator: $P_{C}^{A}P_{B}^{C}=P_{B}^{A}.$

The components $n_{A}$ satisfy the following relations%
\begin{equation}
n^{A}e_{A}^{\alpha }=0,\text{ \ }n_{A}n^{A}=\varepsilon .  \label{16a}
\end{equation}%
To prove this let us note that it follows from (\ref{13}) that $\mathbf{v}%
_{A}\cdot \mathbf{n}=\varepsilon n_{A}.$ Substituting here the expansion 
\begin{equation}
\mathbf{n}=\tilde{n}^{B}\mathbf{v}_{B},  \label{16b}
\end{equation}%
we infer that $\tilde{n}^{B}=\varepsilon n^{B}$ and hence%
\begin{equation}
\mathbf{n}=\varepsilon n^{B}\mathbf{v}_{B}=\varepsilon \left(
n^{B}e_{B}^{\alpha }\mathbf{e}_{\alpha }+n^{B}n_{B}\mathbf{n}\right) ,
\label{16c}
\end{equation}%
from which (\ref{16a}) immediately follows.

In vielbein formalism the soldering form $e_{A}^{\alpha }$ is a fundamental
quantity and the group of symmetry under which the theory is required to be
invariant is the group of \textit{local} Lorentz transformations (\ref{1a}),
where $\Lambda _{A}^{\quad B}=\Lambda _{A}^{\quad B}\left( x\right) .$ Under
Lorentz transformation we have%
\begin{equation}
\mathbf{\tilde{v}}_{A}=\Lambda _{A}^{\hspace{0.05in}B}\mathbf{v}_{B}=\Lambda
_{A}^{\hspace{0.05in}B}\left( e_{B}^{\alpha }\mathbf{e}_{\alpha }+n_{B}%
\mathbf{n}\right) =\tilde{e}_{A}^{\alpha }\mathbf{e}_{\alpha }+\tilde{n}_{A}%
\mathbf{n,}  \label{17}
\end{equation}%
and hence%
\begin{equation}
e_{A}^{\alpha }\rightarrow \tilde{e}_{A}^{\alpha }=\Lambda _{A}^{\hspace{%
0.05in}B}e_{B}^{\alpha }  \label{18}
\end{equation}%
The transformation law for the spin connection follows from its definition:%
\begin{equation*}
\tilde{\omega}_{\beta A}^{\quad \,B}\mathbf{\tilde{v}}_{B}=-\mathbf{\nabla }%
_{\beta }\mathbf{\tilde{v}}_{A}
\end{equation*}%
Substituting $\mathbf{\tilde{v}}_{B}=\Lambda _{A}^{\quad C}\mathbf{v}_{C}$
and taking into account (\ref{5a}) we infer that 
\begin{equation}
\omega _{\mu A}^{\quad B}\rightarrow \tilde{\omega}_{\mu A}^{\quad \hspace{%
0.03in}B}=\left( \Lambda \omega _{\mu }\Lambda ^{-1}\right)
_{A}^{\,B}+\left( \Lambda \partial _{\mu }\Lambda ^{-1}\right) _{A}^{\,B},
\label{20}
\end{equation}%
where $\Lambda $ and $\Lambda ^{-1}$ are the matrices corresponding to
Lorentz transformation and its inverse. Up to this point, we have considered
only vector representations of the Lorentz group. In general, 
\begin{equation}
\Lambda =\exp \left( \lambda ^{AB}J_{AB}\right)
\end{equation}%
where $J_{AB}$ are corresponding generators of the Lie algebra which satisfy
the commutation relations%
\begin{equation}
\left[ J_{AB},J_{CD}\right] =\frac{1}{2}\left( \eta _{BC}J_{AD}-\eta
_{AC}J_{BD}-\eta _{BD}J_{AC}+\eta _{AD}J_{BC}\right)
\end{equation}%
Consider spinors $\psi $ which transforms according to 
\begin{equation}
\psi \rightarrow \exp \left( \frac{1}{4}\lambda ^{AB}\Gamma _{AB}\right)
\psi ,  \label{21}
\end{equation}%
where$\ \Gamma _{AB}=\frac{1}{2}\left( \Gamma _{A}\Gamma _{B}-\Gamma
_{B}\Gamma _{A}\right) $ are generators of the Lie algebra in the spinor
representation and $\Gamma _{A}$ are $d+1$ Dirac matrices satisfying%
\begin{equation}
\left\{ \Gamma ^{A},\Gamma ^{B}\right\} =2\eta ^{AB},\quad \Gamma ^{\dagger
A}=\Gamma ^{0}\Gamma ^{A}\Gamma ^{0}.  \label{22}
\end{equation}%
We note that the signature of $\eta ^{AB}$ does not play any significant
role in the derivations that follow, and thus our results holds equally well
for both de Sitter and anti de Sitter tangent groups. The Dirac action 
\begin{equation}
\dint d^{4}x\sqrt{g}\,\overline{\psi }i\Gamma ^{C}e_{C}^{\alpha }D_{\alpha
}\psi ,  \label{23}
\end{equation}%
where \ 
\begin{equation}
D_{\alpha }\equiv \partial _{\alpha }+\frac{1}{4}\omega _{\alpha }^{\hspace{%
0.03in}AB}\Gamma _{AB},  \label{24}
\end{equation}%
is invariant under gauge transformations (\ref{18}), (\ref{20}) and (\ref{21}%
). This action is real, thanks to the metricity conditions (\ref{11}).

Next one constructs the curvature of the connection $D_{\mu }$ defined by%
\begin{equation}
\left[ D_{\mu },D_{\nu }\right] =\frac{1}{4}R_{\mu \nu }^{\hspace{0.05in}%
\hspace{0.05in}AB}\Gamma _{AB},  \label{25}
\end{equation}%
where%
\begin{equation}
R_{\mu \nu }^{\hspace{0.05in}\hspace{0.05in}AB}\left( \omega \right)
=\partial _{\mu }\omega _{\nu }^{\,\,\,AB}-\partial _{\nu }\omega _{\mu
}^{\,\,\,AB}+\omega _{\mu }^{\,\,\,AC}\omega _{\nu C}^{\quad B}-\omega _{\nu
}^{\,\,\,AC}\omega _{\mu C}^{\quad B}.  \label{26}
\end{equation}%
This curvature$\ $transforms as 
\begin{equation}
\left( R_{\mu \nu }\right) _{A}^{\hspace{0.05in}B}\rightarrow \left( \Lambda
R\Lambda ^{\,-1}\right) _{A}^{\hspace{0.05in}B},  \label{27}
\end{equation}%
and hence 
\begin{equation}
\,R\left( \omega \right) =\,e_{A}^{\mu }R_{\mu \nu }^{\hspace{0.05in}\hspace{%
0.05in}AB}\left( \omega \right) e_{B}^{\nu },  \label{28}
\end{equation}%
is invariant under local gauge transformations. The gauge invariant action
is then given by 
\begin{equation}
S=-\frac{1}{2\kappa ^{2}}\dint d^{4}x\sqrt{g}R\left( \omega \right)
\label{29}
\end{equation}%
Although this action appears to depend on the non-diagonal $e_{A}^{\mu }$,
it is a function of $g_{\mu \nu }$ only.

To prove this we first find how the tangent space covariant derivative acts
on the components of a vector $\mathbf{l=}l^{C}\mathbf{v}_{C}.$ Using spinor
representation for the vector we have%
\begin{equation}
D_{\nu }\left( l^{D}\Gamma _{D}\right) =\partial _{\nu }l^{D}\Gamma _{D}+%
\frac{1}{4}\omega _{\nu }^{\hspace{0.05in}BC}\left[ \Gamma _{BC},\Gamma _{D}%
\right] l^{D}.  \label{30}
\end{equation}%
Taking into account the commutation relation $\left[ \Gamma _{BC},\Gamma _{D}%
\right] =2\left( \eta _{CD}\Gamma _{B}-\eta _{BD}\Gamma _{C}\right) $ one
gets%
\begin{equation}
D_{\nu }\left( l^{D}\Gamma _{D}\right) =\left( \partial _{\nu }l^{D}+\omega
_{\nu \hspace{0.05in}C}^{\hspace{0.05in}D}l^{C}\right) \Gamma _{D},
\label{31}
\end{equation}%
and hence we deduce%
\begin{equation}
D_{\nu }l^{D}=\partial _{\nu }l^{D}+\omega _{\nu \hspace{0.05in}C}^{\hspace{%
0.05in}D}l^{C}.  \label{32}
\end{equation}%
In particular, it follows that 
\begin{equation}
D_{\nu }e^{\rho A}=\partial _{\nu }e^{\rho A}+\omega _{\nu \hspace{0.05in}%
B}^{\hspace{0.05in}A}e^{\rho B},  \label{33}
\end{equation}%
which in turn implies that 
\begin{equation}
\left[ D_{\mu },D_{\nu }\right] e^{\rho A}=R_{\mu \nu }^{\hspace{0.03in}%
\hspace{0.05in}AB}\left( \omega \right) e_{B}^{\rho }.  \label{34}
\end{equation}%
On the other hand, using metricity condition (\ref{11}), we have%
\begin{equation}
D_{\nu }e^{\rho A}=-\Gamma _{\nu \sigma }^{\rho }e^{\sigma A},  \label{35}
\end{equation}%
and therefore 
\begin{align}
D_{\mu }\left( D_{\nu }e^{\rho A}\right) & =-D_{\mu }\left( \Gamma _{\nu
\sigma }^{\rho }e^{\sigma A}\right) =-\left( \partial _{\mu }\Gamma _{\nu
\sigma }^{\rho }\right) e^{\sigma A}-\Gamma _{\nu \sigma }^{\rho }\left(
D_{\mu }e^{\sigma A}\right)  \notag \\
& =-\partial _{\mu }\Gamma _{\nu \sigma }^{\rho }e^{\sigma A}+\Gamma _{\nu
\sigma }^{\rho }\Gamma _{\mu \kappa }^{\sigma }e^{\kappa A}.  \label{36}
\end{align}%
Taking the commutator one gets 
\begin{align}
\left[ D_{\mu },D_{\nu }\right] e^{\rho A}& =-\left( \partial _{\mu }\Gamma
_{\nu \sigma }^{\rho }-\partial _{\nu }\Gamma _{\mu \sigma }^{\rho }+\Gamma
_{\mu \kappa }^{\rho }\Gamma _{\nu \sigma }^{\kappa }-\Gamma _{\nu \kappa
}^{\rho }\Gamma _{\mu \sigma }^{\kappa }\right) e^{\sigma A}  \notag \\
& =-R_{\,\,\,\sigma \mu \nu }^{\rho }\left( \Gamma \right) e^{\sigma A}.
\label{37}
\end{align}%
Comparing this result with (\ref{34}) we arrive at the identity 
\begin{equation}
R_{\mu \nu }^{\hspace{0.05in}\hspace{0.05in}AB}\left( \omega \right)
e_{B}^{\rho }=-R_{\,\,\,\sigma \mu \nu }^{\rho }\left( \Gamma \right)
e^{\sigma A},  \label{38}
\end{equation}%
which in turn leads to 
\begin{align}
\,R\left( \omega \right) & =e_{A}^{\mu }R_{\mu \nu }^{\hspace{0.05in}\hspace{%
0.05in}AB}\left( \omega \right) e_{B}^{\nu }=-R_{\,\,\,\sigma \mu \nu }^{\nu
}\left( \Gamma \right) e^{\sigma A}e_{A}^{\mu }  \notag \\
& =R_{\,\,\,\sigma \nu \mu }^{\nu }\left( \Gamma \right) g^{\sigma \mu
}=R\left( \Gamma \right) .  \label{39}
\end{align}%
This completes the proof that the gauge invariant action (\ref{29}) is
equivalent to Einstein action and involves only those combinations of $%
e_{A}^{\mu }$ which reduce to the metric $g_{\mu \nu }$. The remaining $%
\frac{1}{2}d\left( d+1\right) $ independent combinations of $e_{A}^{\mu }$
components represent the $\frac{1}{2}d\left( d+1\right) $ gauge degrees of
freedom associated with $SO(1,d)$. Thus, we conclude that it is possible to
formulate Einstein gravity as a gauge invariant theory with the tangent
group being de Sitter or anti de Sitter.

We would like to stress that in proving identity (\ref{39}) we never (and
could not) assume that the soldering form $e_{A}^{\mu }$ has an inverse and,
moreover, this result is valid for an arbitrary dimension of tangent space.
However, as it was noticed above the theory is well defined only if $N=d$ or 
$N=d+1.$ We could also consider a gauge invariant action involving higher
order curvature invariants. One can show that even in this case the action
depends only on the metric $g_{\mu \nu }.$ To give an example consider all
possible terms which are of second order in curvature 
\begin{equation}
R_{\mu \nu }^{\hspace{0.05in}\hspace{0.05in}AB}R_{\rho \sigma }^{\hspace{%
0.05in}\hspace{0.05in}CD}\left( c_{1}\ e_{A}^{\mu }e_{B}^{\nu }e_{C}^{\rho
}e_{D}^{\sigma }+c_{2}\ e_{A}^{\mu }e_{C}^{\nu }e_{D}^{\rho }e_{B}^{\sigma
}+c_{3}\ e_{C}^{\mu }e_{D}^{\nu }e_{A}^{\rho }e_{B}^{\sigma }\right) ,\ 
\end{equation}%
because other terms are related to these three by symmetry. The first term
is identical to $R^{2}\left( \Gamma \right) $, while for the second term we
have 
\begin{equation}
R_{\mu \nu }^{\hspace{0.05in}\hspace{0.05in}AB}\left( \omega \right)
e_{A}^{\mu }e_{B}^{\sigma }R_{\rho \sigma }^{\hspace{0.05in}\hspace{0.05in}%
CD}\left( \omega \right) e_{C}^{\nu }e_{D}^{\rho }=g^{\mu \kappa
}R_{\,\,\,\kappa \mu \nu }^{\sigma }\left( \Gamma \right) g^{\nu \lambda
}R_{\,\,\,\lambda \rho \sigma }^{\rho }\left( \Gamma \right) .
\end{equation}%
after using the identity (\ref{38}) twice. Similarly, the third term gives 
\begin{equation}
R_{\mu \nu }^{\hspace{0.05in}\hspace{0.05in}AB}\left( \omega \right)
e_{A}^{\rho }e_{B}^{\sigma }R_{\rho \sigma }^{\hspace{0.05in}\hspace{0.05in}%
CD}\left( \omega \right) e_{C}^{\mu }e_{D}^{\nu }=g^{\kappa \rho
}R_{\,\,\,\kappa \mu \nu }^{\sigma }\left( \Gamma \right) g^{\mu \lambda
}R_{\,\,\,\lambda \rho \sigma }^{\nu }\left( \Gamma \right) ,
\end{equation}%
which proves that the most general action which is second order in
spin-connection curvature is identical to the one that depends on
affine-connection curvature.

\section{Matter couplings}

We have seen that gravity is insensitive to the gauge group of the tangent
space. In this section we will show that, to the contrary, matter
\textquotedblleft feels\textquotedblright\ the tangent space group. Let us
consider the matter couplings in the case of de Sitter tangent group. In
this case the fundamental spinors, vectors and tensors are defined as
representations of the 5d Lorentz group of tangent space, and their
Lagrangians must be invariant with respect to de Sitter symmetry. In
vierbein formulation of gravity, we can exchange space-time tensors with
Lorentz tensors. This is no longer valid for de Sitter tangent group because
in this case the vielbein $e_{A}^{\mu }$ is not invertible and, for example,
a vector in the tangent space is not equivalent to a space-time vector. In
fact as we will show now the 5d de Sitter vector is equivalent to 4d space
time vector and real space time scalar. Therefore, de Sitter tangent space
\textquotedblleft unifies\textquotedblright\ 4d vectors and scalars.

Let us consider a 5d vector $\mathbf{H,}$ which can be expanded in terms of
components as (see (\ref{13}), (\ref{14})): 
\begin{equation}
\mathbf{H}=H^{A}\mathbf{v}_{A}=H^{A}e_{A}^{\alpha }\mathbf{e}_{\alpha
}+H^{A}n_{A}\mathbf{n=}H^{\alpha }\mathbf{e}_{\alpha }+\phi \mathbf{n,}
\label{40a}
\end{equation}%
where 
\begin{equation}
H^{\alpha }=H^{A}e_{A}^{\alpha },\text{ \ }\phi =H^{A}n_{A},  \label{41a}
\end{equation}%
are the components of a 4d vector and a scalar, respectively. Multiplying
the first equation by $e_{\alpha }^{B}$ and taking into account (\ref{16})
we derive 
\begin{equation}
H^{B}=H^{\alpha }e_{\alpha }^{B}+\varepsilon \phi n^{B};  \label{42a}
\end{equation}%
since $e_{\alpha }^{B}n_{B}=0$ and $n^{A}n_{A}=\varepsilon $ (see (\ref{16a}%
)) it follows from here that 
\begin{equation}
H^{B}H_{B}=g_{\alpha \beta }H^{\alpha }H^{\beta }+\varepsilon \phi ^{2}.
\label{43aa}
\end{equation}%
Let us construct the curvature of $H_{A}$%
\begin{equation}
F_{AB}=D_{A}H_{B}-D_{B}H_{A},  \label{44a}
\end{equation}%
where $D_{A}\equiv e_{A}^{\alpha }D_{\alpha }$ and $D_{\alpha }$ is
covariant derivative with respect to tangent space vector indices (see (\ref%
{32})); therefore, the components with only space time indices are scalars
with respect to this derivative, for example, $D_{\alpha }H^{\beta
}=\partial _{\alpha }H^{\beta }.$ Taking this into account and using
decomposition (\ref{42a}) we find%
\begin{equation}
D_{A}H^{B}=e_{A}^{\beta }e_{\alpha }^{B}\partial _{\beta }H^{\alpha
}+e_{A}^{\beta }H^{\alpha }D_{\beta }e_{\alpha }^{B}+\varepsilon
e_{A}^{\beta }n^{B}\partial _{\beta }\phi +\varepsilon e_{A}^{\beta }\phi
D_{\beta }n^{B}.  \label{45a}
\end{equation}%
The last term here is equal to zero. In fact, using the definition (\ref{5a}%
) we have 
\begin{equation}
\partial _{\beta }n_{A}=\varepsilon \mathbf{\nabla }_{\beta }\left( \mathbf{v%
}_{A}\cdot \mathbf{n}\right) =-\omega _{\beta A}^{\quad \
B}n_{B}+\varepsilon \mathbf{v}_{A}\cdot \mathbf{\nabla }_{\beta }\mathbf{n,}
\label{46aaa}
\end{equation}%
and hence $D_{\beta }n_{A}=-\varepsilon \mathbf{v}_{A}\cdot \mathbf{\nabla }%
_{\beta }\mathbf{n.}$ In turn, one can immediately conclude from $\mathbf{%
\nabla }_{\beta }\left( \mathbf{e}_{\alpha }\cdot \mathbf{n}\right) =0$ and $%
\mathbf{\nabla }_{\beta }\left( \mathbf{n}\cdot \mathbf{n}\right) =0$ that $%
\mathbf{\nabla }_{\beta }\mathbf{n=}0$ and therefore $D_{\beta }n_{A}=0.$
Using metricity condition (\ref{35}) to express $D_{\beta }e_{\alpha }^{B}$
in terms of $\Gamma _{\nu \sigma }^{\rho }$ and interchanging indices we
then find%
\begin{equation}
F_{AB}=e_{A}^{\beta }e_{B}^{\alpha }\left( \partial _{\beta }H_{\alpha
}-\partial _{\alpha }H_{\beta }\right) +\varepsilon \left( e_{A}^{\beta
}n_{B}-e_{B}^{\beta }n_{A}\right) \partial _{\beta }\phi .  \label{47a}
\end{equation}%
Note that $F_{AB}$ is invariant under the $U(1)$ gauge transformation%
\begin{equation}
H_{A}\rightarrow H_{A}+e_{A}^{\alpha }\partial _{\alpha }\Lambda ,
\label{48a}
\end{equation}%
which in terms of the space time components become $H_{\alpha }\rightarrow
H_{\alpha }+\partial _{\alpha }\Lambda ,$ $\phi \rightarrow \phi .$ Squaring
(\ref{47a}) we will find the gauge invariant Lagrangian density for the
massless vector field 
\begin{equation}
L=-\frac{1}{4}F_{AB}F^{AB}=-\frac{1}{4}F_{\alpha \beta }F^{\alpha \beta }-%
\frac{1}{2}\varepsilon \partial _{\alpha }\phi \partial ^{\alpha }\phi ,
\label{49a}
\end{equation}%
where 
\begin{equation}
F_{\alpha \beta }=\partial _{\alpha }H_{\beta }-\partial _{\beta }H_{\alpha
}.  \label{50a}
\end{equation}%
Notice that we get the correct sign for the kinetic energy of the scalar
field $\phi $ only in the case of de Sitter group ($\varepsilon =-1$) while
for anti de Sitter group $\varepsilon =1$ we get a ghost$.$ We deduce that
the formulation of gravity where the tangent group is $SO(1,d)$ instead of $%
SO(1,d-1)$ unifies spins zero and spin one in one vector field. If we add to
the Lagrangian the term (\ref{43aa}) both fields acquire the same mass.

We now turn to spinors. Because they should respect 5d tangent Lorentz group
it is well known that neither Majorana or Weyl conditions can be imposed on
them \cite{Scherk}. Thus the spinors $\psi $ must be Dirac spinors. The
Dirac action in this case is 
\begin{equation*}
\dint \sqrt{g}d^{4}x\left( i\overline{\psi }\Gamma ^{A}D_{A}\psi -i\overline{%
D_{A}\psi }\Gamma ^{A}\psi \right)
\end{equation*}%
The spinors do feel the full $SO(1,4)$ local symmetry. This seems to be a
very strong constraint as it implies that chiral spinors cannot exist if the
tangent group is $SO(1,4).$ This is similar to the situation in case of
supersymmetry in five dimensions \cite{Cremmer}, \cite{CN}, or $N=2$
supersymmetry. There, it was shown that it is possible to generalize the
Majorana condition by taking a doublet of spinors \cite{Scherk}. The
conclusion we must draw is then that the $SO(1,4)$ tangent group implies
that spinors must be treated in the same way as in $N=2$ supersymmetry. To
couple the spinors to vectors, some gauge symmetry must be introduced. As an
example, let us assume the existence of a $U(1)$ gauge symmetry. In this
case the covariant derivative $D_{A}\psi $ becomes 
\begin{equation}
D_{A}\psi =\left( e_{A}^{\mu }\left( \partial _{\mu }+\frac{1}{4}\omega
_{\mu }^{\hspace{0.03in}AB}\Gamma _{AB}\right) \ +iH_{A}\right) \psi \ ,
\end{equation}%
which shows that the spinors exist in a unified interactions with both a
scalar and a vector field, as was seen in the decomposition of the vector $%
H_{A}$ into a vector $H_{\mu }$ and a scalar $\phi .$

\section{Complex gravity and unitary $U(1,d-1)$ tangent group}

As a tangent space one can also consider a complex vector space with
Hermitian scalar product satisfying 
\begin{equation}
\left( \mathbf{v},\mathbf{u}\right) =\left( \mathbf{u},\mathbf{v}\right)
^{\ast },\text{ \ }\left( \mathbf{v},\alpha \mathbf{u}\right) =\alpha \left( 
\mathbf{v},\mathbf{u}\right) ,  \label{40}
\end{equation}%
where $\alpha $ is a complex number. It follows from here that $\left(
\alpha \mathbf{v},\mathbf{u}\right) =\alpha ^{\ast }\left( \mathbf{v},%
\mathbf{u}\right) .$ As before let us introduce in this space the
orthonormal basis $\mathbf{v}_{A}$ $(A=1,...N)$: 
\begin{equation}
\left( \mathbf{v}_{A},\mathbf{v}_{B}\right) =\eta _{AB}.  \label{41}
\end{equation}%
The condition of orthogonality is preserved under $U(1,N-1)$ transformations%
\begin{equation}
\mathbf{\tilde{v}}_{A}=U_{A}^{\hspace{0.05in}C}\mathbf{v}_{C},\ \ \ \ \ \
U_{A}^{\hspace{0.05in}C}\eta _{CD}\left( U_{A}^{\hspace{0.05in}D}\right)
^{\ast }=\eta _{AB}.  \label{43}
\end{equation}%
For generality let us first consider the complex coordinate basis vectors $%
\mathbf{e}_{\alpha }$ $(\alpha =1,...d)$ in $d$-dimensional manifold and
show that in this case we obtain the Hermitian theory of gravity as
formulated by Einstein and Strauss \cite{Ein}, \cite{ES}. Later on we will
show that this theory can be consistently truncated to General Relativity
while preserving the unitary structure of the tangent space.

Assuming that $N\geq d$ \ we can expand the coordinate basis vectors in
terms of vielbein vectors, $\mathbf{e}_{\alpha }=e_{\alpha }^{A}\mathbf{v}%
_{A},$ and then the metric on the manifold can be expressed as%
\begin{equation}
g_{\alpha \beta }\equiv \left( \mathbf{e}_{\alpha },\mathbf{e}_{\beta
}\right) =e_{\alpha }^{A}e_{\beta }^{B\ast }\eta _{AB}.  \label{44}
\end{equation}%
This metric is Hermitian%
\begin{equation*}
g_{\alpha \beta }=\left( \mathbf{e}_{\alpha },\mathbf{e}_{\beta }\right)
=\left( \mathbf{e}_{\beta },\mathbf{e}_{\alpha }\right) ^{\ast }=g_{\beta
\alpha }^{\ast }.
\end{equation*}%
In the case under consideration the affine and spin connections are defined
exactly as in (\ref{5a}). Taking derivative of (\ref{44}) and using
definition in (\ref{5a}) we obtain%
\begin{equation}
\partial _{\gamma }g_{\alpha \beta }=\left( \mathbf{\nabla }_{\gamma }%
\mathbf{e}_{\alpha },\mathbf{e}_{\beta }\right) +\left( \mathbf{e}_{\alpha },%
\mathbf{\nabla }_{\gamma }\mathbf{e}_{\beta }\right) =\Gamma _{\alpha \gamma
}^{\nu \ast }g_{\nu \beta }+\Gamma _{\beta \gamma }^{\nu }g_{\alpha \nu }.
\label{45}
\end{equation}%
These $d^{\,3}$ equations can be solved unambiguously for $\Gamma _{\kappa
\rho }^{\mu }$ in terms of metric $g_{\alpha \beta }$ only if we impose the
hermiticity condition%
\begin{equation}
\Gamma _{\rho \mu }^{\nu \ast }=\Gamma _{\mu \rho }^{\nu },  \label{46}
\end{equation}%
which leaves us with $d^{3}$ components to be determined. Unlike the real
case equations (\ref{46}) can be solved only perturbatively. They were first
imposed by Einstein in his formulation of Hermitian gravity which he
referred to as the "$+-$" condition \cite{Ein}, \cite{ES}, \cite{DD}.
Similar to (\ref{6}) we derive a condition on spin connection%
\begin{equation}
\omega _{\alpha A}^{\quad C}\eta _{CB}=-\left( \omega _{\alpha B}^{\quad
C}\right) ^{\ast }\eta _{CA},  \label{46a}
\end{equation}%
which leaves $N^{2}d$ independent components. Taking derivative of $\left( 
\mathbf{v}_{A},\mathbf{e}_{\alpha }\right) =e_{\alpha }^{B}\eta _{AB}$ we
derive the following metricity conditions 
\begin{equation}
\partial _{\gamma }e_{\alpha }^{A}=\omega _{\gamma B}^{\quad \,A}e_{\alpha
}^{B}+\Gamma _{\alpha \gamma }^{\nu }e_{\nu }^{A}.  \label{47}
\end{equation}%
Taking into account that $d^{3}$ equations (\ref{45}) determine $\Gamma
_{\beta \gamma }^{\nu }$ through $\partial _{\gamma }e_{\alpha }^{A}$ we are
left with $2Nd^{2}-d^{3}$ equations to find $N^{2}d$ independent components
of $\omega _{\alpha A}^{\quad C}.$ The number of equations match the number
of unknown components only if $N=d,$ that is, when dimension of complex
tangent space coincides with the dimension of the manifold. Hence the gauge
group of the tangent space can be only $U(1,d-1)$ \cite{AHC}. In this case
we can define the soldering form $e_{B}^{\beta },$ which is inverse to $%
e_{\alpha }^{A}:$%
\begin{equation}
e_{B}^{\alpha }e_{\alpha }^{A}=\delta _{B}^{A},\text{ \ \ }e_{A}^{\alpha
}e_{\beta }^{A}=\delta _{\beta }^{\alpha }.\text{\ }  \label{48}
\end{equation}%
The metric with upper indices is then given by%
\begin{equation}
g^{\mu \nu }=e_{A}^{\mu }e_{B}^{\nu \ast }\eta ^{AB},  \label{49}
\end{equation}%
and it is inverse to $g_{\alpha \beta }$ 
\begin{equation}
g_{\alpha \nu }g^{\beta \nu }=\delta _{\alpha }^{\beta }\neq g_{\alpha \nu
}g^{\nu \beta }.  \label{50}
\end{equation}%
Similar to (\ref{34}) the curvature of the connection $\omega _{\mu
A}^{\quad B}$ can be defined as 
\begin{align}
\left[ D_{\mu },D_{\nu }\right] e_{A}^{\sigma }& \equiv R_{\mu \nu A}^{\quad 
\hspace{0.03in}\hspace{0.03in}B}\left( \omega \right) e_{B}^{\sigma }  \notag
\\
& =\left( \partial _{\mu }\omega _{\nu A}^{\quad B}-\partial _{\nu }\omega
_{\mu A}^{\quad B}+\omega _{\mu A}^{\quad C}\omega _{\nu C}^{\quad B}-\omega
_{\nu A}^{\quad C}\omega _{\mu C}^{\quad B}\right) e_{B}^{\sigma }.
\label{52}
\end{align}%
On the other hand, using the metricity condition, we have 
\begin{align}
\left[ D_{\mu },D_{\nu }\right] e_{A}^{\sigma }& =-\left( \partial _{\mu
}\Gamma _{\rho \nu }^{\sigma }-\partial _{\nu }\Gamma _{\rho \mu }^{\sigma
}+\Gamma _{\kappa \mu }^{\sigma }\Gamma _{\rho \nu }^{\kappa }-\Gamma
_{\kappa \nu }^{\sigma }\Gamma _{\rho \mu }^{\kappa }\right) e_{A}^{\sigma }
\notag \\
& \equiv -R_{\hspace{0.03in}\hspace{0.03in}\rho \mu \nu }^{\sigma }\left(
\Gamma \right) e_{A}^{\rho },  \label{53}
\end{align}%
and it follows from here that%
\begin{equation}
R_{\hspace{0.03in}\hspace{0.03in}\rho \mu \nu }^{\sigma }\left( \Gamma
\right) =-e_{\rho }^{A}R_{\mu \nu A}^{\quad \,\,\,\,B}\left( \omega \right)
e_{B}^{\sigma }.  \label{54}
\end{equation}%
In particular, the scalar curvature 
\begin{align}
R\left( \omega \right) & =\eta ^{AC}e_{C}^{\mu \ast }R_{\mu \nu A}^{\quad
\,\,\,\,B}\left( \omega \right) e_{B}^{\nu }=-\eta ^{AC}e_{C}^{\mu \ast
}R_{\,\,\,\rho \mu \nu }^{\nu }\left( \Gamma \right) e_{A}^{\rho }  \notag \\
& =g^{\rho \mu }R_{\,\,\,\rho \nu \mu }^{\nu }\left( \Gamma \right) =R\left(
\Gamma \right) ,  \label{55}
\end{align}%
is $U\left( 1,d-1\right) $ gauge invariant. The scalar curvature is real,%
\begin{equation}
R^{\ast }\left( \omega \right) =R\left( \omega \right) .  \label{56}
\end{equation}%
\ To prove this we first note the identity%
\begin{equation}
\left( R_{\mu \nu A}^{\quad \,\,\,B}\left( \omega \right) \right) ^{\ast
}=-R_{\mu \nu C}^{\quad \,\,\,\,D}\left( \omega \right) \eta ^{CB}\eta _{DA},
\label{57}
\end{equation}%
which follows from equation (\ref{52}) taking into account (\ref{46a}).
Using this relation together with (\ref{54}) we obtain 
\begin{align}
\left( R_{\hspace{0.03in}\hspace{0.03in}\rho \mu \nu }^{\sigma }\left(
\Gamma \right) \right) ^{\ast }& =-e_{\rho }^{A\ast }\left( R_{\mu \nu
A}^{\quad \,\,\,B}\left( \omega \right) \right) ^{\ast }e_{B}^{\sigma \ast
}=e_{\rho }^{A\ast }R_{\mu \nu C}^{\quad \,\,\,D}\left( \omega \right) \eta
^{CB}\eta _{DA}e_{B}^{\sigma \ast }  \notag \\
& =-\eta ^{CB}e_{C}^{\kappa }e_{B}^{\sigma \ast }R_{\,\,\,\kappa \mu \nu
}^{\lambda }\left( \Gamma \right) \eta _{DA}e_{\lambda }^{D}e_{\rho }^{A\ast
}=-g^{\kappa \sigma }R_{\,\,\,\kappa \mu \nu }^{\lambda }\left( \Gamma
\right) g_{\lambda \rho }.  \label{58}
\end{align}%
It follows from here that the tensor 
\begin{equation}
R_{\rho \kappa \mu \nu }\left( \Gamma \right) =R_{\,\,\,\kappa \mu \nu
}^{\lambda }\left( \Gamma \right) g_{\lambda \rho },  \label{59}
\end{equation}%
is antihermitian with respect to exchange of first two indices 
\begin{equation}
\left( R_{\kappa \rho \mu \nu }\left( \Gamma \right) \right) ^{\ast
}=-R_{\rho \kappa \mu \nu }\left( \Gamma \right) ,  \label{60}
\end{equation}%
and it is antisymmetric with respect to exchange of the last two indices
(see (\ref{53}). Taking this into account we have%
\begin{equation}
R^{\ast }\left( \Gamma \right) =\left( g^{\rho \mu }g^{\nu \sigma }R_{\sigma
\rho \nu \mu }\right) ^{\ast }=g^{\mu \rho }g^{\sigma \nu }R_{\rho \sigma
\mu \nu }=R\left( \Gamma \right) ,  \label{61}
\end{equation}%
and because $R\left( \omega \right) =R\left( \Gamma \right) ,$ this
completes the proof of reality of gauge invariant scalar curvature.

The identity (\ref{60}) was not noticed by Einstein and this forced him to
construct Hermitian combinations of the curvature tensor. As we see this is
not necessary because one can use instead the real scalar curvature as
Lagrangian density.

If we write the connection as 
\begin{equation}
\omega _{\mu A}^{\quad B}=\bar{\omega}_{\mu A}^{\quad B}+\frac{1}{d}\hat{%
\omega}_{\mu }\delta _{A}^{B},  \label{62}
\end{equation}%
where 
\begin{equation}
\bar{\omega}_{\mu A}^{\quad A}=0,\text{ \ }\hat{\omega}_{\mu }=\omega _{\mu
A}^{\quad A},  \label{63}
\end{equation}%
the curvature splits into two pieces 
\begin{equation}
R_{\mu \nu A}^{\quad \,\,\,B}\left( \omega \right) =R_{\mu \nu A}^{\quad
\,\,B}\left( \bar{\omega}\right) +\frac{1}{d}R_{\mu \nu C}^{\quad
\,\,\,C}\left( \hat{\omega}\right) \delta _{A}^{B},  \label{64}
\end{equation}%
where 
\begin{align}
R_{\mu \nu A}^{\quad \,\,\,B}\left( \bar{\omega}\right) & =\left( \partial
_{\mu }\bar{\omega}_{\nu A}^{\quad B}-\partial _{\nu }\bar{\omega}_{\mu
A}^{\quad B}+\bar{\omega}_{\mu A}^{\quad C}\bar{\omega}_{\nu C}^{\ \ \ \ B}-%
\bar{\omega}_{\nu A}^{\quad C}\bar{\omega}_{\mu C}^{\ \ \ \ B}\right) , 
\notag \\
R_{\mu \nu C}^{\quad \,\,\,C}\left( \omega \right) & =\partial _{\mu }\hat{%
\omega}_{\nu }-\partial _{\nu }\hat{\omega}_{\mu }.  \label{65}
\end{align}%
It follows from here that 
\begin{equation}
R\left( \omega \right) =\eta ^{AC}e_{C}^{\mu \ast }R_{\mu \nu A}^{\quad
\,\,\,B}\left( \bar{\omega}\right) e_{B}^{\nu }+\frac{1}{d}g^{\nu \mu
}R_{\mu \nu C}^{\quad \,\,\,C}\left( \hat{\omega}\right) =R\left( \bar{\omega%
}\right) +\frac{1}{d}\tilde{R}\left( \hat{\omega}\right) ,  \label{66}
\end{equation}%
where $\tilde{R}=g^{\nu \mu }R_{\mu \nu A}^{\quad \,\,\,A}$ is another
scalar curvature invariant. \ Therefore it can be added to the action with
an arbitrary coefficient leading to the following most general gauge
invariant first order action%
\begin{equation}
S=\dint d^{4}x\left\vert \det e_{\mu }^{A}\right\vert \left( \alpha R\left( 
\bar{\omega}\right) +\beta \tilde{R}\left( \hat{\omega}\right) \right) .
\label{67}
\end{equation}%
It must be stressed that we are using here a second order formalism where
the field $\omega _{\mu A}^{\quad B}$ is determined by the metricity
condition and not by the field equations. The best strategy to analyze this
action is to solve for $\omega _{\mu A}^{\quad B}$ in a perturbative
expansion in terms of $e_{\mu }^{A}.$

We can understand the above results by noting that the gauge invariant
action allows to use the gauge invariance to reduce the independent
components of $e_{\alpha }^{A}$ to those of $g_{\alpha \beta }.$ In other
words we expect that because of $U(1,d-1)$\ gauge invariance, the action
depends only on the metric%
\begin{equation*}
g_{\alpha \beta }=e_{\alpha }^{A}e_{\beta }^{B\ast }\eta _{AB}\equiv
G_{\alpha \beta }+iB_{\alpha \beta }.
\end{equation*}%
This theory was considered before using a first order formalism where the
spin-connection was determined from the equations of motion \cite{AHC}%
\textit{. }This is possible only when the action depends quadratically on
the spin-connection. However, the $U(1)$ part $\hat{\omega}$ of the $%
U(1,d-1) $ connection being abelian, appears linearly. This then imposes a
constraint on the antisymmetric part of the metric%
\begin{equation}
\partial _{\alpha }\left( \left\vert \det e_{\mu }^{A}\right\vert B^{\alpha
\beta }\right) =0,
\end{equation}%
which thus remains undetermined \cite{AHC}. This is to be contrasted with
the second order formalism where all spin-connections are determined from
the metricity condition. \textit{\ }

We arrive to an interesting case by requiring that the metric $g_{\alpha
\beta }$ to be real. This is equivalent to truncating the $B_{\alpha \beta }$
field. Let 
\begin{equation}
e_{\alpha }^{A}=e_{\alpha \left( 0\right) }^{A}+ie_{\alpha \left( 1\right)
}^{A},
\end{equation}%
so that 
\begin{eqnarray}
G_{\alpha \beta } &=&\left( e_{\alpha \left( 0\right) }^{A}e_{\beta \left(
0\right) }^{B}+e_{\alpha \left( 1\right) }^{A}e_{\beta \left( 1\right)
}^{B}\right) \ \eta _{AB}, \\
B_{\alpha \beta } &=&\left( e_{\alpha \left( 1\right) }^{A}e_{\beta \left(
0\right) }^{B}-e_{\alpha \left( 0\right) }^{A}e_{\beta \left( 1\right)
}^{B}\right) \ \eta _{AB},
\end{eqnarray}%
Truncating $B_{\alpha \beta }$ gives $\frac{1}{2}d\left( d-1\right) $
constraints on the $2d^{2}$ (real) fields $e_{\alpha \left( 0\right) }^{A}$
and $e_{\alpha \left( 1\right) }^{A}$. In this case the affine connection is
also real and its $\frac{1}{2}d^{2}\left( d+1\right) $ components are
Christoffel connection for the metric $G_{\alpha \beta }.$ The remaining%
\begin{equation*}
2d^{3}-\frac{1}{2}dd\left( d-1\right) -\frac{1}{2}d^{2}\left( d+1\right)
=d^{3}
\end{equation*}%
independent equations (\ref{47}) are then enough to unambiguously determine $%
d^{3}$ components of $\omega _{\mu A}^{\quad B}.$ This implies that it is
possible to enlarge the tangent group to become $U(1,d-1)$ and still obtain
the Einstein gravity without any modification. The coupling to matter will,
however, feel the tangent group $U(1,d-1)$.

\textbf{Matter coupling.} When the tangent group is $U(1,3)$ then from the
previous discussion it should be clear that neither the Majorana nor the
Weyl condition could be imposed, except if a doublet of spinors is taken.
Thus, as with the $SO(1,4)$ case we must take a Dirac spinor, or a doublet
of Majorana or Weyl spinors, again as in the $N=2$ supersymmetric case. We
note the isomorphism of the algebras 
\begin{equation}
U\left( 1,3\right) \sim SO(1,5)\times SO(1,1).
\end{equation}%
It is easy to see that $U\left( 1,3\right) $ has ten compact generators and
six non-compact generators, while $SO(1,5)$ has ten compact generators and
five non-compact generators and $SO(1,1)$ has one non-compact generator.
Thus spinors in the case of unitary tangent group will exhibit conformal
local symmetry.

Gravity has a universal coupling to matter. One way to classify the fields
is according to their behavior under the diffeomorphism group, or
equivalently under the tangent Lorentz group. A complex scalar field has the
following couplings%
\begin{equation}
\dint d^{4}x\sqrt{\det g}g^{\mu \nu }\partial _{\mu }\phi \partial _{\nu
}\phi ^{\ast }.
\end{equation}%
For a massless vector it can be easily seen that the action can be written
in terms of a complex space-time vector $H_{\mu }$\ with the action 
\begin{equation}
\ \dint d^{4}x\sqrt{\det g}g^{\mu \rho }g^{\nu \sigma }F_{\mu \nu }F_{\rho
\sigma }^{\ast }.
\end{equation}%
Similarly we can treat the case of fields which are in the vector
representations of the gauge group. The fermions have more complicated
couplings. First, a Dirac spinor has the $U(1,d-1)$ transformation%
\begin{equation}
\psi \rightarrow e^{i\lambda _{B}^{\hspace{0.03in}A}\Gamma _{A}\Gamma
^{B}}\psi ,
\end{equation}%
where $\Gamma ^{A}$ and $\Gamma _{A}$ satisfy the relations 
\begin{equation*}
\left\{ \Gamma ^{A},\Gamma ^{B}\right\} =0,\quad \left\{ \Gamma _{A},\Gamma
_{B}\right\} =0,\quad \left\{ \Gamma ^{A},\Gamma _{B}\right\} =\delta
_{B}^{A},
\end{equation*}%
and thus $\Gamma _{A}\Gamma ^{B}$ are the generators of $U(1,d-1).$ We can
define the Hermitian Dirac matrices 
\begin{align*}
\gamma ^{\mu }& =e_{A}^{\mu }\Gamma ^{A}+e^{\mu A}\Gamma _{A}, \\
\left\{ \gamma ^{\mu },\gamma ^{\nu }\right\} & =g^{\mu \nu }+g^{\nu \mu },
\end{align*}%
The covariant derivative is given by 
\begin{equation*}
D_{\mu }\psi =\partial _{\mu }\psi +\omega _{\mu B}^{\quad A}\Gamma
_{A}\Gamma ^{B}\psi ,
\end{equation*}%
Hermitian Dirac action is then 
\begin{equation}
\dint d^{4}x\left\vert \det e_{\mu }^{A}\right\vert \overline{\psi }\gamma
^{\mu }D_{\mu }\psi .
\end{equation}%
Therefore, Dirac spinors do couple to both the symmetric and antisymmetric
components of the Hermitian metric.

\section{Conclusions}

We have shown that Einstein gravity exhibits universality when formulated as
a gauge theory of tangent space group. Besides of the well known natural
case when the tangent space has the same dimension as the manifold, we
discovered two other possibilities for General Relativity to be reproduced
and the theory still remains unambiguous. Namely, we have shown that in the
four dimensional case the tangent space can be five dimensional and possess
(anti) de Sitter group of symmetry. This group is important when we
incorporate matter couplings to the gravitational field. As an example, we
have shown that de Sitter tangent space group allows us to \textquotedblleft
unify\textquotedblright\ 4d vectors and scalars which become components of
the same five dimensional vector in tangent space. Even more dramatic are
the consequences of the tangent space symmetry group on fermions. They
become fundamentally five dimensional and neither Majorana nor Weyl
conditions could be imposed on them. This situations is similar to $N=2$
supersymmetry where we are forced to generalize the Majorana condition by
taking a doublet of spinors. We also would like to note that if we impose an
extra $U\left( 1\right) $ local symmetry in the tangent space then the
spinors would exist in a unified interaction with both scalar and vector
fields.

Another interesting possibility arise when we consider complex tangent space
of the same dimension as the manifold. In this case the group of symmetry is
the unitary group. This gives rise generically to the theory of Hermitian
gravity, where the basic fields are the symmetric and antisymmetric
components of the metric, which coincide with the basic fields appearing in
effective open string field theory. It is interesting that this theory can
be consistently truncated to Einstein gravity, while still preserving the
unitary group of tangent space. In turn, this has interesting and nontrivial
consequences for the coupling to matter which should respect this symmetry.
In a forthcoming paper \cite{CM} we shall explore the implications of these
new formulations of gravity, especially in regard to the spontaneous
breakdown of these larger symmetries down to the $SO(1,d-1)$ symmetry.

\section{Appendix: The Poincare limit and 3d CS gravity}

In this appendix we examine the special case when the radius of the de
Sitter tangent group becomes infinite, which corresponds to Poincare
symmetry. Later we shall also investigate the correspondence with  
Chern-Simons gravity in three dimensions which also have de Sitter or
Poincare symmetry \cite{AT}, \cite{Witten}.

The $SO(1,d)$ group generators satisfy the commutation relations%
\begin{equation}
\left[ J_{AB},J_{CD}\right] =-\frac{1}{2}\left( \eta _{AC}J_{BD}-\eta
_{BC}J_{AD}-\eta _{AD}J_{BC}+\eta _{BD}J_{AC}\right) .
\end{equation}%
Splitting the range of the index $A=a,\overline{d},$ \ where $a=0,1,\cdots
,d-1,$ and similarly for the other indices we get the usual $SO(1,d-1)$ for
the $J_{ab},$ while for $J_{a\overline{d}}\equiv RP_{a}$ we have 
\begin{equation}
\left[ P_{a},P_{b}\right] =-\frac{1}{R^{2}}J_{ab}.
\end{equation}%
Thus, in the limit $R\rightarrow \infty $ the de Sitter tangent group
becomes the inhomogeneous Lorentz group, i.e. $ISO(1,d-1)$ also known as the
Poincare group. The covariant derivative 
\begin{equation}
D_{\mu }=\partial _{\mu }+\omega _{\mu }^{\,\,AB}J_{AB},
\end{equation}%
implies that the field $\omega _{\mu }^{a\overline{d}}$ must be defined as $%
\omega _{\mu }^{a\overline{d}}\equiv \frac{1}{2R}b_{\mu }^{a}$ so that 
\begin{equation}
D_{\mu }=\partial _{\mu }+\omega _{\mu }^{\,\,ab}J_{ab}+b_{\mu }^{\,a}P_{a},
\end{equation}%
is independent of the radius $R.$ The curvatures in terms of the redefined
fields are 
\begin{align}
R_{\mu \nu }^{\quad ab}& =\partial _{\mu }\omega _{\nu }^{\,\,\,ab}-\partial
_{\nu }\omega _{\mu }^{\,\,\,ab}+\omega _{\mu }^{\,\,\,ac}\omega _{\nu
c}^{\quad b}-\omega _{\nu }^{\,\,\,ac}\omega _{\mu c}^{\quad b}-\frac{1}{%
4R^{2}}\left( b_{\mu }^{\,a}b_{\nu }^{\,b}-b_{\mu }^{\,b}b_{\nu
}^{\,a}\right) , \\
R_{\mu \nu }^{\quad a\overline{d}}& =\frac{1}{2R}\left( \partial _{\mu
}b_{\nu }^{\,a}-\partial _{\mu }b_{\nu }^{\,a}+\omega _{\mu
}^{\,\,\,ac}b_{\nu c}-\omega _{\nu }^{\,\,\,ac}b_{\mu c}\right) .
\end{align}%
The zero torsion condition on $e_{A}^{\mu }$ is consistent in the limit $%
R\rightarrow \infty $ if we define 
\begin{equation}
e_{\overline{d}}^{\mu }\equiv \frac{1}{R}c^{\mu },
\end{equation}%
so that 
\begin{equation}
\ \partial _{\mu }c^{\nu }-\frac{1}{2}b_{\mu }^{\,a}e_{a}^{\nu }+\Gamma
_{\rho \mu }^{\nu }c^{\rho }=0,
\end{equation}%
which allows us to calculate $b_{\mu }^{a}$ in terms of $c^{\mu }.$ The
field $\omega _{\mu }^{\quad ab}$ is solved from the condition 
\begin{equation}
\ \partial _{\mu }e_{a}^{\nu }+\omega _{\mu a}^{\quad b}e_{b}^{\nu }+\frac{1%
}{2R^{2}}b_{\mu }^{a}c^{\nu }+\Gamma _{\rho \mu }^{\nu }e_{a}^{\rho }=0.
\end{equation}%
Writing the gravitational action in terms of the rescaled fields, we expand $%
e_{A}^{\mu }R_{\mu \nu }^{\hspace{0.05in}\hspace{0.05in}AB}\left( \omega
\right) e_{B}^{\nu }$ to get 
\begin{align}
& e_{a}^{\mu }e_{b}^{\nu }\left( \partial _{\mu }\omega _{\nu }^{\hspace{%
0.05in}ab}-\partial _{\nu }\omega _{\mu }^{\,\hspace{0.05in}ab}+\omega _{\mu
}^{\,\,\,ac}\omega _{\nu c}^{\quad b}-\omega _{\nu }^{\,\,\,ac}\omega _{\mu
c}^{\quad b}-\frac{1}{4R^{2}}\ \left( b_{\mu }^{\,a}b_{\nu }^{\,b}-b_{\mu
}^{\,b}b_{\nu }^{\,a}\right) \right)  \notag \\
& \ +\frac{1}{R^{2}}e_{a}^{\mu }c^{\nu }\left( \partial _{\mu }b_{\nu
}^{\,a}-\partial _{\nu }b_{\mu }^{\,a}+\omega _{\mu }^{\hspace{0.05in}%
ac}b_{\nu c}-\omega _{\nu }^{\hspace{0.05in}ac}b_{\mu c}\right) .
\end{align}%
Therefore it is clear that in the limit $R\rightarrow \infty $ the
connection $\omega _{\mu }^{\,\,ab}$ coincides with the $SO(1,d-1)$ Lorentz
connection and the action becomes identical to the Einstein-Hilbert action.
The fields $b_{\mu }^{\,a}$ and $c^{\mu }$ drop out of the action. Thus in
the limit of $ISO(1,d-1)$ the action is indistinguishable from the $%
SO(1,d-1) $ invariant action for gravity.

For matter couplings, especially for the vector $H_{A}$, the gauge
transformation is 
\begin{equation*}
\delta H_{A}=\lambda _{AB}H^{B},\quad \lambda _{AB}=-\lambda _{BA}.
\end{equation*}%
Denoting $H_{\overline{d}}=\phi $ \ and $\lambda _{a\overline{d}}=\frac{1}{2R%
}\lambda _{a}$, the gauge transformations of $H_{a}$ and $\phi $ are 
\begin{align*}
\delta H_{a}& =\lambda _{ab}H^{b}+\frac{1}{2R}\lambda _{a}\phi , \\
\delta \phi & =-\frac{1}{2R}\lambda _{a}H^{a}.
\end{align*}%
Thus, in the limit $R\rightarrow \infty $ the fields $H_{a}$ and $\phi $
remain in the action as spin one and spin zero fields, but they decouple in
the transformations and become independent.

When our $SO(1,d)$ gauge invariant gravitational action is taken in three
dimensions, it is natural to ask whether the action obtained is identical to
the Chern-Simons action which was also shown by Achucarro-Townsend \cite{AT}
and Witten \cite{Witten} to be equivalent to the Einstein action in three
dimensions, but with a cosmological constant. In the Chern-Simons
construction one uses only the gauge field $\omega _{\mu }^{AB}$ where the
CS action is 
\begin{equation}
I_{\mathrm{CS}}=\frac{1}{2}\dint d^{3}x\epsilon ^{\mu \nu \rho }\epsilon
_{ABCD}\left( \omega _{\mu }^{\,\,AB}\partial _{\nu }\omega _{\rho
}^{\,\,CD}+\frac{2}{3}\omega _{\mu }^{\,\,AB}\omega _{\nu }^{\,\,CE}\omega
_{\rho E}^{\quad \,D}\right) .
\end{equation}%
Using the same decomposition for $\omega _{\mu }^{AB}$ as before, we get 
\begin{equation}
\frac{1}{R}\dint d^{3}x\epsilon ^{\mu \nu \rho }\epsilon _{abc}b_{\mu
}^{a}\left( \partial _{\nu }\omega _{\rho }^{\,\,bc}+\omega _{\nu
}^{\,\,be}\omega _{\rho e}^{\quad c}-\frac{1}{12R^{2}}b_{\nu }^{\,b}b_{\rho
}^{\,\,c}\right) ,
\end{equation}%
which is the the first order formulation of the Einstein action plus a
cosmological constant, with the dreibein field $b_{\mu }^{\,a}$. The special
case with the $ISO(1,d-1)$ gauge group can be recovered by rescaling the
action by $R$ and then taking the limit $R\rightarrow \infty .$ In our
treatment, there is also the additional field $e_{A}^{\mu }$ which is not a
gauge field. The field $b_{\mu }^{\,a}$ is given by 
\begin{equation*}
b_{\mu }^{\,a}=2e_{\nu }^{a}\nabla _{\mu }c^{\nu },
\end{equation*}%
where $e_{\nu }^{a}$ is the inverse of $e_{a}^{\nu }.$ Our action can be
expressed in terms of $e_{a}^{\nu }$ and a non-propagating field $c^{\mu }.$
Comparing the two formulations, we deduce that the field $b_{\mu }^{\,a}$
must be identified with $e_{\mu }^{a}$. Although $e_{\mu }^{a}$ is not a
gauge field, it can be shown, using the torsion constraint, that its
diffeomorphism transformation with parameters $\zeta ^{\mu }$ can yield the
same gauge transformation as $b_{\mu }^{\,a}$ with the gauge parameter $%
\lambda ^{a}=e_{\mu }^{a}\zeta ^{\mu }$ \cite{Witten}. It then clear that
although both formulations have the same gauge symmetry, they have different
field configurations. Moreover, the usual matter couplings in the CS
formulation are not possible because the dreibein $b_{\mu }^{\,a}$ is a
gauge field. Any direct coupling to matter breaks gauge invariance, except
for coupling to Wilson lines. In our case since $e_{A}^{\mu }$ is not a
gauge field, a gauge invariant metric can be easily formed $g^{\mu \nu
}=e_{A}^{\mu }e^{\nu A}$ and coupled to any form of matter desired.

\begin{acknowledgement}
The work of AHC is supported in part by the Alexander von Humboldt
Foundation and by the National Science Foundation 0854779. V.M. is supported
by TRR 33 \textquotedblleft The Dark Universe\textquotedblright\ and the
Cluster of Excellence EXC 153 \textquotedblleft Origin and Structure of the
Universe\textquotedblright .
\end{acknowledgement}

\end{document}